# Detecting self-similarity in surface microstructures

R. Piasecki

*Institute of Chemistry, University of Opole, Oleska 48, PL 45052 Opole, Poland*

*E-mail:* piaser@uni.opole.pl

**Abstract**

The relative configurational entropy per cell as a function of length scale is a sensitive detector of spatial self-similarity. For Sierpinski carpets the equally separated peaks of the above function appear at the length scales that depend on the kind of the carpet. These peaks point to the presence of self-similarity even for randomly perturbed initial fractal sets. This is also demonstrated for the model population of particles diffusing over the surface considered by Van Siclen, Phys. Rev. E 56 (1997) 5211. These results allow the subtle self-similarity traces to be explored.

**1. Introduction**

In this article we examine the length scale behaviour of the finite-sized object (FSO) extension [1] of the recently introduced entropic measure [2] of spatial inhomogeneity. The nonuniform planar surfaces in adsorption processes or the irregular internal surfaces in porous materials are frequently modelled by diluted self-similar sets. The notion of self-similarity is widely employed when the geometric irregularities in materials are statistically length scale independent. For example, the recent experimental study of multilayer adsorption on fractal surfaces in porous media [3], the surface reaction monomer-monomer model on fractal lattices [4] and the influence of lacunarity of regular fractal lattices on the dynamic scaling behaviour of a monomer-dimer model [5], all fall into this group.

The (qualitative) evaluation of microstructure attributes enables a search for their possible connections with macroscopic properties of a medium. In our case, the traces of self-similarity in disordered planar configurations of FSOs can be detected by a simple method proposed in this paper. By objects we understand the black pixels of binary image obtained from an experiment or, as in this case, generated by a computer. In order to test the method, randomly perturbed Sierpinski carpets, population of particles diffusing over the surface and lacunary fractal lattices were chosen. To quantify the degree of spatial inhomogeneity we use the recently discussed relative configurational entropy per cell for FSOs [1]. Our method can be also used to three-dimensional data. However, such experimental data is not so readily available, see for instance [6,7] where an approach [8] related to the information entropy has been applied.

## 2. Entropic measure description

In previous papers [1,2] full descriptions of the entropic measure were given, where $S_\Delta$ and $f(S)$ refer to the case of FSOs and 'point' objects (POs), respectively. Although the general concept of the two measures is the same, we present the short derivation of $S_\Delta$ indicating the essential differences. For a given $L \times L$ binary image, let $0 < n < L^2$ of the black pixels be distributed in square and nonoverlapping lattice cells of size $k \times k$. Obviously, in the PO approach the number of objects is not restricted. At every length scale $k$ equal to the actual cell-side length and commensurate with $L$, we have $\chi = (L/k)^2$ of the lattice cells. Besides the standard constraint for the cell occupation numbers, $n_1 + ... + n_\chi = n$, there is now one more: $n_i \leq k^2$ for each $i = 1, 2,..., \chi$. Some specific configurations possible for POs, e.g. when all of them are placed in the same cell, cannot be realized by FSOs.

The simplest approach is to consider all the distributions of black pixels with fixed occupation numbers as a kind of scale-dependent configurational macrostate described by the set $\{n_i\}$. Indeed, for a given length scale $k$ any macrostate can be realized by a number of distinguishable arrangements of $n$ black pixels associated with the $1 \times 1$ lattice cells, i.e. some kind of equally likely configurational microstates with $n_i = 0, 1$. Setting the Boltzmann constant $k_B = 1$ we can use a standard definition of configurational entropy: $S(k, L, n) = \ln \Omega(k, L, n)$, where the number of the appropriate microstates is given by

$$\Omega(k,L,n) = \prod_{i=1}^{\chi} \binom{k^2}{n_i}. \tag{1}$$

For each length scale $1 \leq k \leq L$ the highest possible value of configurational entropy $S_{max}(k, L, n) = \ln \Omega_{max}(k, L, n)$ is related to the most uniformly distributed object configuration. Furthermore such a configuration represents a so-called reference configurational macrostate $\{n_i\}_{RCM}$ described by the following condition: for each pair $i \neq j$ we have $|n_i - n_j| \leq 1$. Thus, for a given $\{n_i\}_{RCM}$ the maximal number of the proper configurational microstates equals

$$\Omega_{max}(k, L, n) = \binom{k^2}{n_0}^{\chi - r_0} \binom{k^2}{n_0 + 1}^{r_0}, \qquad (2)$$

where $r_0 = n \bmod \chi$, $r_0 \in \{0, 1, ..., \chi - 1\}$ and $n_0 = (n - r_0)/\chi$, $n_0 \in \{0, 1, ..., k^2 - 1\}$.

For a given image we shall concentrate on the dependence of the entropic measure $S_\Delta$ of spatial inhomogeneity on the length scale $k$. To evaluate for each $k$ the deviation of the actual configuration from the appropriate $\{n_i\}_{RCM}$ it is natural to consider the difference $S_{max} - S$. Averaging this difference over the number of cells $\chi$ we obtain a highly sensitive spatial object arrangement measure $S_\Delta(k, L, n) \equiv [S_{max} - S]/\chi$. This averaging is necessary to obtain (see below) the crucial property of the measure which allows for its calculation at every length scale. Taking into account Eqs.(1) and (2) and the definition of $S_\Delta$, the final form of the measure can be written as follows

$$S_\Delta(k, L, n) = \frac{r_0}{\chi} \ln\left(\frac{k^2 - n_0}{n_0 + 1}\right) + \frac{1}{\chi} \sum_{i=1}^{\chi} \ln\left[\binom{k^2}{n_i} \bigg/ \binom{k^2}{n_0}\right]. \qquad (3)$$

The measure $S_\Delta(k, L, n)$ exhibits a number of simple properties:

1. The lowest possible value is equal to 0 and is always reached at boundary length scales, i.e. $k = 1$ and $k = L$, while $S_\Delta(0 < k < L) = 0$ shows that a configuration belongs to the maximally ordered macrostate.

2. The highest value for a given $k$ is reached when fully occupied and/or empty cells and at most one cell partially filled appear. Such a configuration represents a possible maximally disordered macrostate corresponding to the strongest deviation per cell from an appropriate $\{n_i\}_{RCM}$.

3. For a given $k$ the first well-shaped peak of $S_\Delta(k, L, n)$ indicates that the cells strongly

occupied $n_0 \ll n_i$ as well as weakly filled $n_i \ll n_0$ dominate. We can safely say that the tendency to clustering of objects is marked in this way. The repeated decreasing maxima indicate the clustering of clusters. For Sierpinski carpets the sequential peaks of various heights are equally separated.

4. The deep minima in $S_\Delta(k, L, n)$ describe relatively more ordered configuration where the main contribution comes from the cells occupied by $n_i \approx n_0$ particles. The sequential and equally separated, distinct minima reflect periodicity of the whole configuration.

5. For binary images with equal number $n$ of objects the height of maxima and depth of minima allow for the comparison of relative intensity of deviations per cell from an appropriate $\{n_i\}_{\text{RCM}}$.

6. The following property allows us to calculate the value of the measure at *every* length scale, i.e. for $k = 1, 2, 3,..., L$. If a final pattern of size $mL \times mL$ (where $m$ is a natural number) is formed by periodic repetition of an initial arrangement of size $L \times L$, then the value of the measure at a given length scale $1 \leq k \leq L$ commensurate with the side length $L$ is *unchanged* under the replacement $L \leftrightarrow mL$, since it also causes $n \leftrightarrow m^2 n$, $\chi \leftrightarrow m^2 \chi$, $r_0 \leftrightarrow m^2 r_0$ keeping $n_0$ and the corresponding $n_i$ the same. To overcome the problem of an incommensurate length scale it is sufficient to determine a number $m'$ such that $m'L \bmod k = 0$ and replace the initial arrangement of size $L \times L$ by the periodically created one of size $m'L \times m'L$. Then define $S_\Delta(k, L, n) \equiv S_\Delta(k, m'L, m'^2 n)$.

7. The relation for a homogeneous function of the second degree in length variables $k$ and $L$ is fulfilled, that is $S_\Delta(\lambda k, \lambda L) = \lambda^\alpha S_\Delta(k, L)$ with $\alpha \approx 2$ for a planar configuration. This property extends also to other dimensions and will be discussed elsewhere.

8. One more interesting the 'symmetry' property of the measure is not dependent on the length scale $k$. Namely, the measure value does not change under the replacement of 'black phase' for 'white phase' and vice versa, that is $S_\Delta(k, L, n) = S_\Delta(k, L, n')$, where $n + n' = L^2$. So, for the fraction $n/L^2$ of black pixels the spatial disorder is 'seen' to be the same as for the inverted state with a $1 - n/L^2$ concentration of white pixels.

## 3. Model structures and concluding remarks

For square Sierpinski carpets the distinctive behaviour of $S_\Delta$ can be observed [1] at different length scales. Namely, the sequential peaks in $S_\Delta$ are separated by an interval of constant length while for other non-fractal random sets the peaks are placed irregularly. Such a behaviour of $S_\Delta$ can be utilized to detect the self-similarity in binary images of complex microstructures or perturbed fractal sets. For deterministic and random Sierpinski carpets we use the notation DSC(a, b, c) and RSC(a, b, c). The meaning of parameters *a*, *b* and c results from the simple construction procedure. After dividing an initial square lattice of $L \times L$ cells with $L = a^c$ into $a^2$ subsquares only *b* of them are conserved according to deterministic rule or at random. This segmentation is repeated on each conserved subsquare, and so on, *c* times. By definition a Sierpinski carpet has the fractal dimension $d_f = \ln b / \ln a$. Let us consider a few different illustrative examples.

(I)   For the standard DSC(3, 8, 3) with $d_f \approx 1{,}89$ the characteristic behaviour of $S_\Delta$ as a function of length scale *k* is presented in Fig. 1 (solid line). The inset shows the configuration with the identical number and size distribution of 'white' squares (empty places) like for the standard DSC(3, 8, 3) but now the positions of the squares are generated sequentially and randomly beginning from the largest one. The corresponding values of $S_\Delta$ (black circles) run in somewhat different way in comparison with the solid line but the small peaks for $k = 9, 18, 27, ..., 72$ are still present. Thus the traces of 'self-similar on average' microstructure or equivalently 'statistical self-similarity' are seen to be detected in this way. Such a behaviour implies the appearance of similar proportional distributions of occupation numbers for the increasing length scales but not necessarily the same local ordering.

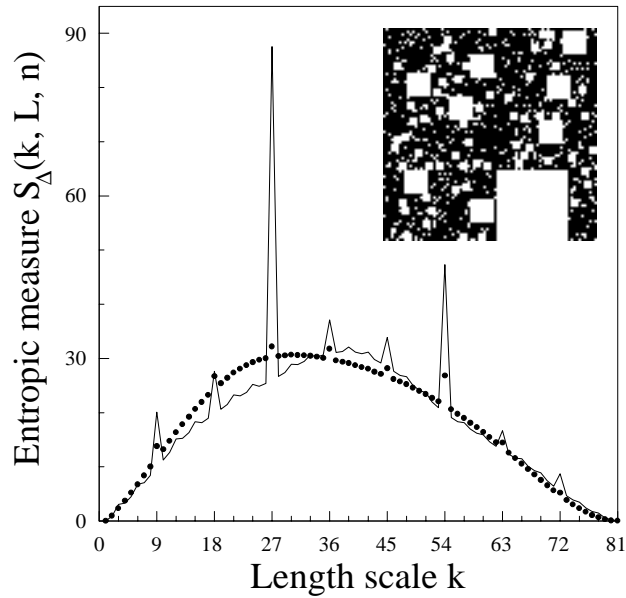

**Fig. 1.** $S_\Delta(k, L, n)$ vs. *k* for the standard DSC(3, 8, 4) with $L = 81$ and $n = 4096$ (solid line) and connected randomly perturbed spatial arrangement showed in the inset (filled circles).

(II) The upper right inset in Fig. 2 shows another initial fractal RSC(5, 16, 3) with $d_f \approx 1{,}72$. The related two sets perturbed in a different way than before, i.e. by the random subtraction and addition of black pixels, are shown in the upper left and lower insets with 6% and 50% black pixel fractions. Despite the strong deviations from the initial fraction (26,21%) the two additional $S_\Delta$ curves imitate the behaviour of the curve for the initial random fractal surprisingly well. The equally separated and well-marked peaks for $k = 25, 50, 75$ and $100$ clearly indicate the statistical self-similarity. As expected, both subtraction and addition of the objects suppress the spatial inhomogeneity at every length scale.

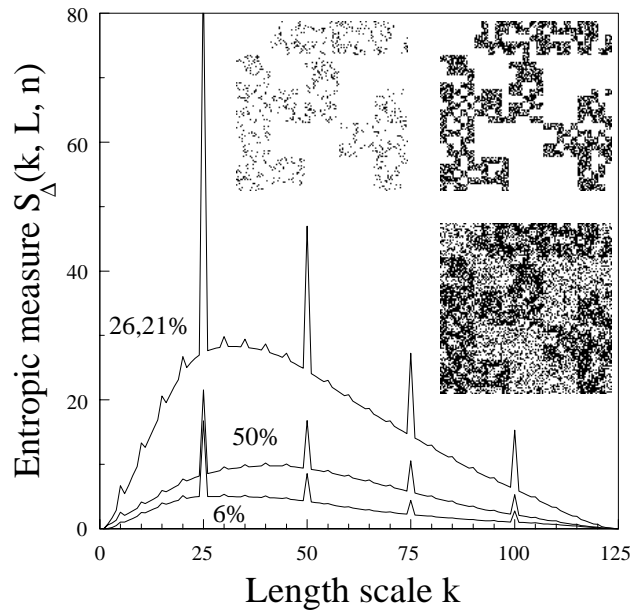

**Fig. 2.** $S_\Delta(k, L, n)$ vs. $k$ for an example of RSC(5, 16, 3) with $L = 125$ and $n = 4096$ (upper right inset) and two connected sets perturbed by the random subtraction of 3158 (upper left inset) and addition of 3717 (lower inset) black pixels. The related fractions are also indicated.

(III) The same kind of perturbation is now applied to a population of interacting particles taken from [9] (cf. Fig. 1). For the particle diffusing randomly over the surface small ramified clusters are formed due to a small binding energy between adjacent particles (the upper right inset with the 25% fraction in Fig. 3). For the two randomly perturbed configurations (upper left and lower insets) the middle-length scales $k = 15$, 20, ..., 40 still reveal the appearance of the sequential peaks. Such a behaviour is similar to that in Sierpinski carpets with a parameter $a = 5$. The simplicity and well-known properties of Sierpinski carpets allow for identification of those peaks with the self-similarity features of the fractal carpets at different length scales. According to the construction of the measure $S_\Delta$ the only reason for such a behaviour in the population under consideration is the appearance of the traces of statistical self-similarity. It provides a simple way of detecting subtle self-similarity traces in inhomogeneous microstructures.

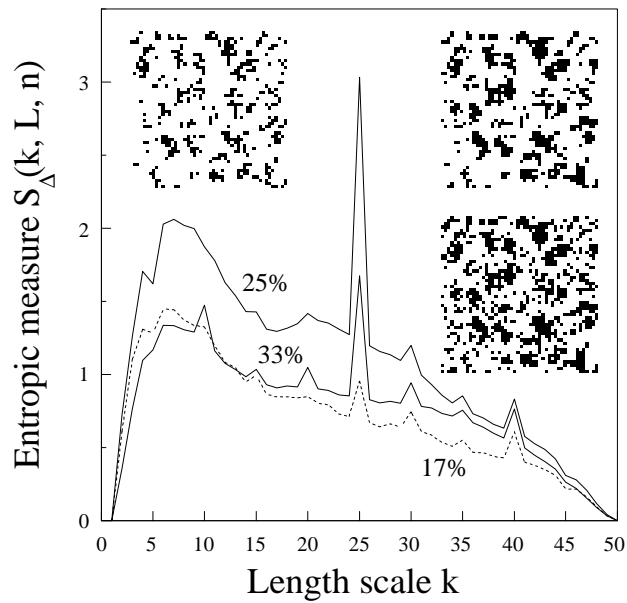

**Fig. 3.** A similar situation as in Fig. 2, but with (adapted from Ref. [9]) set of $n = 625$ interacting particles on linear size $L = 50$ grid (upper right inset, solid line, the 25% fraction) and two connected sets perturbed by the random subtraction (upper left inset, dashed line, 17%) and addition (lower inset, solid line, 33%) of 200 black pixels.

(IV) Fig. 4 shows the plot of $S_\Delta$ for the two Sierpinski fractal lattices with different lacunarity indexes $l_1 = 0.1374$ for the upper left and $l_2 = 0.0556$ for the right inset [5] but the same fractal dimension $d_f \approx 1{,}79$. Such lattices can be used as models for catalytic surfaces. The lattice sites of size $2 \times 2$ in pixels are considered to be located in the centre of the $4 \times 4$ subsquares. The upper $S_\Delta$ curve (solid line) corresponding to $l_1$ lattice and the lower curve (dashed line) describing $l_2$ lattice clearly distinguish the two lattices at most of the length scales except the peculiar scales to these fractal lattices where $S_\Delta(l_1) = S_\Delta(l_2)$. Since the two fractal lattices have the same number of black pixels (or equivalently sites) we can safely say that in this case for all the scales different from the peculiar ones the surface inhomogeneity quantified by $S_\Delta$ is smaller for the lattice with the lower lacunarity index. The behaviour of $S_\Delta$ is in agreement with the meaning of lacunarity, which measures the deviation of a fractal from being translationally invariant.

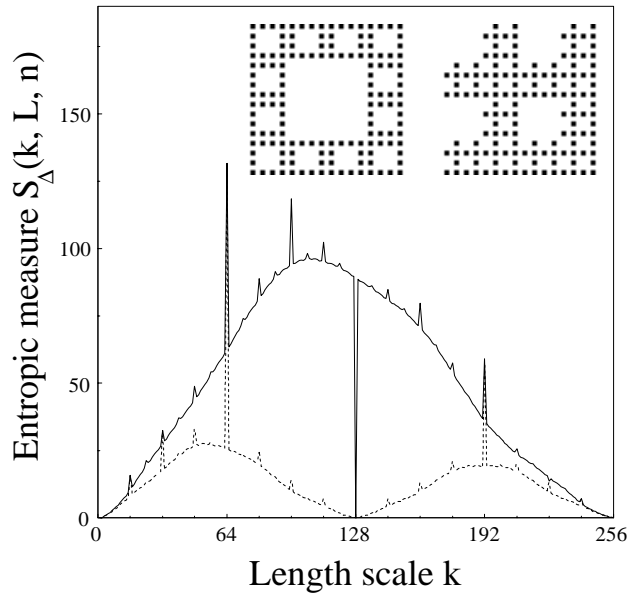

**Fig. 4.** $S_\Delta(k, L, n)$ vs. $k$ for two examples of DSC(4, 12, 3) with $L = 256$ in pixels and $n = 6912$ black pixels or equivalently lattice size of 64 and the number of 1728 sites indicated by $2 \times 2$ black squares (the insets show only the second fragmentation stage). This example adapted from Ref. [5] refers to the case when two fractal lattices have the same fractal dimension $d_f \approx 1{,}79$ but different lacunarities, $l_1 = 0.1374 > l_2 = 0.0556$ for the left and right lattice, respectively.

Concluding, we have analysed using the entropic measure $S_\Delta$ four examples of model surface structures, subsequently perturbed in various ways. For all of the examples the measure detects even weak traces of statistical self-similarity. Examples (I)-(III) also show that statistical self-similarity is a quantity that is hardly affected by various random perturbations. Additionally, example (IV) shows that the lower index of lacunarity *l*, the lower average spatial inhomogeneity. This suggests a possible connection between the two quantities.